\documentclass[final,5p,times,twocolumn]{elsarticle}

\usepackage{amsmath}
\usepackage{mathtools}
\usepackage{graphicx}
\usepackage{ulem}
\usepackage{footnote}
\usepackage{morefloats}
\usepackage{rotating}
\usepackage{relsize}
\usepackage{color}
\usepackage{floatflt,epsfig}
\usepackage{ulem}
\usepackage{natbib}
\usepackage{hyperref}
\usepackage{filecontents}
\usepackage{widetext}

\journal{Physics Letters B}

\begin{document}

\begin{frontmatter}

\title{Effects of modified dispersion relations on free Fermi gas: equations of state and applications in astrophysics}

\author{Luis C. N. Santos\corref{cor1}\fnref{label1}}
\ead{luis.santos@ufsc.br}
\address{Departamento de F\'isica, Universidade Federal da Para\'iba, \\Caixa Postal 5008, 58059-900, Jo\~ao Pessoa, PB, Brazil}

\author{Cl\'esio E. Mota}
\ead{clesio200915@hotmail.com}
\address{Departamento de F\'isica, CFM, Universidade Federal de Santa Catarina;\\ C.P. 476, CEP 88.040-900, Florian\'opolis, SC, Brasil}

\author{Franciele M. da Silva}
\ead{franmdasilva@gmail.com}
\address{Departamento de F\'isica, CFM, Universidade Federal de Santa Catarina;\\ C.P. 476, CEP 88.040-900, Florian\'opolis, SC, Brasil}

\author{Guilherme Grams}
\ead{grams.guilherme@gmail.com}
\address{Univ. Lyon, Univ. Claude Bernard Lyon 1, CNRS/IN2P3, IP2I Lyon, UMR 5822, F-69622, Villeurbanne, France}

\author{I. P. Lobo}
\ead{iarley\_lobo@fisica.ufpb.br}
\address{Department of Chemistry and Physics, Federal University of Para\'iba, Rodovia BR 079 - Km 12, 58397-000 Areia-PB,  Brazil}
\address{Physics Department, Federal University of Lavras, Caixa Postal 3037, 37200-000 Lavras-MG, Brazil}

\fntext[label1]{Corresponding author}

\begin{abstract}
Deformed dispersion relations are considered in the study of equations of state of Fermi gas with applications to compact objects. Different choices of deformed energy relations are used in the formulation of our model. As a first test, we consider a relativistic star with a simple internal structure. The mass-radius diagrams obtained suggest a positive influence of deformed Fermi gas, depending of the functions employed. In addition, we comment on how realistic equations of state, in which interactions between nucleons are taken into account, can be addressed.
\end{abstract}


\end{frontmatter}

\section{Introduction}
\label{sec:intro}

 Despite the many efforts made in recent years, combining gravity with quantum mechanics remains one of  the greatest challenges in theoretical physics. In this context, the study of the Lorentz symmetry at Planck scale plays a important role where the Plank energy $E_p=\sqrt{\hbar c^5 / G}$ acts as a  threshold separating the classical and quantum regimes. At this scale, it is possible to preserve the relativity principle by relying on modifications of special relativity, that can lead to  a deformation  of the Lorentz symmetry. 
Doubly/Deformed special relativity (DSR) theories take into account an invariant energy scale in addition to the invariant velocity scale producing modifications of the dispersion relation with very interesting applications in  astronomical and cosmological observations \cite{magueijo}. Some of these observations, including threshold anomalies in astrophysics data, have been analyzed in \cite{dispersao,dispersao3,dispersao4,dispersao5} considering deformations of special relativity dispersion relation.

An important feature of DSR is the possibility of incorporating effects from different theoretical and observational motivations in a single scheme. For instance, an ultraviolet cut-of predicted by some results in the literature of black holes physics \cite{dispersao6,dispersao7} can be taken into account in a dispersion relation. In addition, the thermodynamics of such a compact object produced by these modifications can induce an uncertainty in the geometrical location of the horizon. \cite{lobo2020effects}. Besides that, effects of uncertainties in the location of the horizon of a black hole due to Planck scale Generalized Uncertainty Principles can be interpreted, in fact, as due to a horizonless compact object, whose phenomenological possibilities have been recently investigated in \cite{Buoninfante:2020cqz}.

In this way, the mentioned consequences suggest a review of possible effects on other compact objects as relativistic stars, where the high density of matter has important effects on the kinetic terms of the equation of state. In order to do that, it is possible to consider a free Fermi gas model in which effects of DSR modify the relation between energy density and pressure in the interior of the star. In this context, the relation between Fermi gases and Lorentz invariance \cite{camacho2006white,gregg2009modified,bertolami2010towards,amelino2012uv,mishra2018invariant}, and  relation in other thermodynamic systems \cite{camacho2007thermodynamics,zhang2011photon,chandra2012thermodynamics,mishra2017equilibrium} have been studied in recent years. Here, we present a modified Fermi gas due to the DSR and study the structure of a simple model of Neutron Star (NS).  The focus of the present work is to investigate the effects of the modified equation of state (EoS) on the mass $\times$ radius diagram.   


In the study of NS it is usual to probe different nuclear interactions, such as Skyrme, Gogny, relativistic mean field (RMF) \cite{Chabanat1997,Lourenco2020,GONZALEZBOQUERA2018195,PhysRevC.82.055803,PhysRevC.99.045202,tm1}, while keeping the kinetic contribution to the energy as a Fermi gas. In the present work we investigate the effect of the DSR on the EoS that describes a Fermi gas without interaction. We analyze the effects of this modification on the macroscopic properties of a simple NS model composed by free neutrons only. In addition to understanding the modification on the kinetic Fermi energy, which is the main objective of the present study, this work is the first step towards a more realistic description of a NS in the context of DSR.

This paper is divided as follows: in section \ref{formal} we introduce the DSR formalism and the modified dispersion relation that we will study in this paper, in section \ref{TOV} we apply our modified EoSs to relativistic stars that can represent a simple model of NS, in section \ref{results} we describe and discuss our results and finally in section \ref{conclusion} we summarise our work and our conclusions.

\section{Formalism} \label{formal}

The formalism that we use here considers a general expression for energy of particles with a broad range of applications. We formulate expressions for the energy density and others quantities of interest by assuming deformations both from theoretical and observational points of view. The low limit energy is investigated, where usual relations from especial relativity are recovered.

\subsection{Dispersion relations}
\label{Rainbowtheory}
The starting point in our model is a
DSR proposal with an invariant energy scale in which the deformation of the usual dispersion relation  $E^2-k^2c^2=m^2 c^4$ can be written as 
\begin{equation}
    E^2 f(x)^{2} - k^2 c^2 g(x)^{2} = m^2 c^4,
    \label{eq0}
\end{equation}
where the argument $x=\lambda E/E_{p}$ involves the ratio of the energy of a test particle to the Planck energy $E_{p}$ with $\lambda$ being a real number (and $k$ is the particle's momentum). This is a general expression to the energy and can be written in a particular form depending on the nature of physical system. In the literature, the functions $f(x)$ and $g(x)$ may be chosen both to reproduce theoretical results expected from the semi-classical limit of quantum gravity and to explain certain phenomenological aspects from astrophysical problems. As a first case, we will analyze the well-known set of functions in which the speed of light remains constant along the physical energy 
\cite{dispersao8} through the choice of functions 
\begin{align}
f(x)=\frac{1}{1-x},\quad g(x)=\frac{1}{1-x}.  
\label{eq1}
\end{align}
As a second case, we will use the following set
\begin{align}
     f(x)=\frac{e^x -1}{x}, \qquad g(x)=1, 
    \label{eq2}
\end{align}
where the exponential form of these functions is useful to explain observational data of gamma ray bursts at cosmological distances \cite{caso2uso1,caso2uso2}.



It is important to note that in the infrared limit the standard energy--momentum dispersion relation is recovered in this type of theory, i.e., the functions $f(x)$ and $g(x)$  satisfy the
conditions:
\begin{equation}
    \lim_{x\rightarrow 0} f(x)=1, \quad \lim_{x \rightarrow 0} g(x)=1.
    \label{eq3}
\end{equation}

\subsection{DSR Fermi Gas}\label{fermi}
Here we examine how changes in the dispersion relation affect a gas of fermions at zero temperature. It is considered $N$ relativistic particles obeying the Fermi--Dirac statistics occupying a volume $V$. Under these assumptions, it turns out the following particle density \cite{dispersao9}
\begin{align}
    dn=f(E)\frac{g}{(2\pi \hbar )^3}d^3k,
    \label{eq4}
\end{align}
where $f(E)=(exp((E-\mu)/k_BT)+1)^{-1}$ is the Fermi--Dirac distribution with $E,\mu,k_B,\hbar$ and $T$ being the energy, chemical potential, Boltzmann constant, Planck constant temperature, respectively. 
Note that at the time of observations the temperature of NS are between $10^{5}$ to $10^{9}$ K, which in the nuclear scale (1 MeV = 1.16 $\times$ $10^{10}$ K) are quite cold. Therefore we take the zero temperature limit in the present study and assume $T=0$ in the Fermi-Dirac distribution \cite{Glendenning1997}. 
In this expression, $g$ gives us the number of states of a particle with a certain value of $k$. We may particularize Eq. (\ref{eq4}) for degenerate fermions assuming $g=2$, and then integrating it, we obtain

\begin{equation}
    n= \int_{0}^{k_f}\frac{8 \pi k^2}{(2 \pi \hbar)^3} dk=\frac{k_{f}^{3}}{3\pi^{2}\hbar^{3}},
    \label{eq5}
\end{equation}
where $k_f$ is the Fermi momentum. With these results it is straightforward to obtain the energy density considering the modified special relativity energy $E_{msr}$ and integrating the relation $d \varepsilon= E_{msr} dn$, with the result being given by

\begin{equation}
    \varepsilon(k_f)=\frac{8 \pi}{(2 \pi \hbar)^3} \int_0^{k_f} E_{msr} k^2 dk.
    \label{eq6}
\end{equation}
In a similar way, the pressure $dp= n (\frac{d}{dk}E_{msr}) dk$ can be written as

\begin{equation}
    p(k_f)=\frac{1}{3 \pi^2 \hbar^3} \int_0^{k_f} \left(\frac{d}{dk}E_{msr}\right) k^3 dk
    \label{eq7}.
\end{equation}
The procedure to determine the final form of Equations (\ref{eq6}) and (\ref{eq7}) requires the use of  relations in Eq. (\ref{eq0}). In our case, we use the relations (\ref{eq1}) and (\ref{eq2}) and then we solve the resulting expression for $E$.   

\subsection{First case}

By substituting Eq. (\ref{eq1}) into Eq. (\ref{eq0}), we get the energy
\begin{equation}
    E_{msr}= A_1 + A_3 \sqrt{k^2 c^2 + \bar{m}^2 c^4},
    \label{eq8}
\end{equation}
where we have defined the parameters
\begin{equation}
    A_1= -\frac{(\lambda/E_p) m^2 c^4 }{1- (\lambda^2 /E_p^2) m^2 c^4 },
    \label{eq9}
\end{equation}

\begin{equation}
    \bar{m}^2= \frac{A_2^2}{A_3^2}=\frac{m^2}{1- (\lambda^2 /E_p^2) m^2 c^4 },
    \label{eq10}
\end{equation}

\begin{equation}
    A_2= \frac{m}{1- (\lambda^2 /E_p^2) m^2 c^4 },
    \label{eq11}
\end{equation}

\begin{equation}
    A_3= \frac{1}{\sqrt{1- (\lambda^2 /E_p^2) m^2 c^4 }}.
    \label{eq12}
\end{equation}
Note that in the limit $\lambda \rightarrow 0$, Eq. (\ref{eq8}) reduces to the usual positive  dispersion relation in special relativity. Indeed, when this consideration it is taken into account, the equations of the modified Fermi gas will match standard results in the literature. So, if one substitutes expression (\ref{eq8}) in Eq. (\ref{eq6}), and then integrates it, the result is the energy density $\varepsilon_{msr}$. In the same way, substituting (\ref{eq8}) in Eq. (\ref{eq7}) we obtain the pressure of the system $p_{msr}$. Thus we write these expressions as    

\begin{widetext}
\begin{equation}
     \varepsilon_{msr}(k_f)= A_1 \frac{8 \pi}{(2 \pi \hbar)^3} \frac{k_f^3}{3} +A_3 \frac{c}{8 \pi^2 \hbar^3} \left\{ k_f \sqrt{k_f^2 + \bar{m}^2 c^2} \left( 2 k_f^2 + \bar{m}^2 c^2 \right) - \bar{m}^4 c^4 \ln \left[ \frac{k_f + \sqrt{k_f^2 + \bar{m}^2 c^2}}{\bar{m} c} \right] \right\},
     \label{eq13}
\end{equation}

\begin{equation}
     p_{msr}(k_f)= A_3 \frac{c}{24 \pi^2 \hbar^3} \left\{ \frac{2 k_f^5 - \bar{m}^2 c^2 k_f^3 -3 \bar{m}^4 c^4 k_f}{\sqrt{k_f^2 + \bar{m}^2 c^2}} + 3 \bar{m}^4 c^4 \ln \left[ \frac{k_f + \sqrt{k_f^2 + \bar{m}^2 c^2}}{\bar{m} c} \right] \right\}
     \label{eq14}.
\end{equation}
\end{widetext}
As it was mentioned above, the equations of our model reduce to the standard special relativity expressions in the appropriate limit. To see this, we observe that in the limit $\lambda \rightarrow 0$ equations (\ref{eq9}), (\ref{eq10}), (\ref{eq11}), and (\ref{eq12}) reduce to the following results respectively
\begin{align}
    A_1=0,\:\:,\bar{m}=m, \:\: A_2=m,\:\:A_3=1,
     \label{eq15}
\end{align}
and, as a consequence, the energy density $\varepsilon_{msr}$ and pressure $p_{msr}$ reduce to the well-known values  in special relativity  $\varepsilon_{sr}$ and $p_{sr}$, respectively. 
Before we analyze the application of these EoSs in the study of NS, we write down the equations of a Fermi gas using the deformed functions in Eq. (\ref{eq2}).

\subsection{Second case}
It is important to study the influence of different deformation functions on the equations of state in order to obtain scenarios that can be applied effectively in astrophysics. For this purpose, we test a second set in which the function $f(x)$ has an exponential form. Now, functions (\ref{eq2}) may used in (\ref{eq0}) and then solved for $E$. The resultant expression, that is related to the positive special relativity dispersion relation, reads as  
\begin{equation}
    E_{msr}=\frac{\ln \left[ 1 + (\lambda/E_p) \sqrt{k^2 c^2 + m^2 c^4} \right]}{(\lambda/E_p)}. 
    \label{eq16}
\end{equation}

Unfortunately, it is difficult to solve and obtain the low energy limit applying this dispersion relation in Eq. (\ref{eq6}) and (\ref{eq7}). The best way to do this is by separating the energy density in a contribution of undeformed special relativity $\varepsilon_{sr}$ and correction terms  $\varepsilon_{msr}$. By applying Eq. (\ref{eq16}) in (\ref{eq6})  and expanding the resulting integral in series (up to second order), we obtain  

\begin{equation}
     \varepsilon(k_f)=\varepsilon_{sr}(k_f)+\varepsilon_{msr}(k_f),
     \label{eq17}
\end{equation}
where we have  
\begin{widetext}
\begin{equation}
     \varepsilon_{sr}(k_f)= \frac{c}{8 \pi ^2 \hbar ^3}\left\{ k_f \sqrt{k_f^2+m^2 c^2} \left(2 k_f^2+m^2 c^2\right) -m^4 c^4 \ln \left[\frac{k_f +\sqrt{k_f^2+m^2 c^2}}{m c}\right]\right\}, 
      \label{eq18}
\end{equation}

\begin{equation}
\begin{split}
     \varepsilon_{msr}(k_f)=&\frac{c}{8 \pi ^2 \hbar ^3}\left\{ k_f \sqrt{k_f^2+m^2 c^2} \frac{\lambda^2 c^2}{18 E_p^2} \right. \left(2 k_f^2 \left( 4 k_f^2+ 7 m^2 c^2\right) +3 m^4 c^4\right) -m^6 c^6 \frac{\lambda^2 c^2}{6 E_p^2} \ln \left[\frac{k_f +\sqrt{k_f^2+m^2 c^2}}{m c}\right] \\
     & \left. -\frac{4 \lambda c}{15 E_p} k_f^3 \left(3 k_f^2+5 m^2 c^2 \right)\right\},  
\end{split}
 \label{eq19}
\end{equation}

\end{widetext}

In a similar form, we can use  (\ref{eq16}) in (\ref{eq7})  and then expand the resulting integral in series, the result is

\begin{equation}
     p(k_f)=p_{sr}(k_f)+p_{msr}(k_f),
      \label{eq20}
\end{equation}

with
\begin{widetext}

\begin{equation}
     p_{sr}(k_f)= \frac{c}{24 \pi^2 \hbar^3} \left\{ \frac{2 k_f^5 - m^2 c^2 k_f^3 -3 m^4 c^4 k_f}{\sqrt{k_f^2 + m^2 c^2}} + 3 m^4 c^4 \ln \left[ \frac{k_f + \sqrt{k_f^2 + m^2 c^2}}{m c} \right] \right\} 
      \label{eq21},
\end{equation}

\begin{equation}
\begin{split}
     p_{msr}(k_f)=& \frac{c}{24 \pi^2 \hbar^3} \left\{ \frac{\lambda^2 c^2}{6 E_p^2} \frac{2 k_f^5 \left( 4 k_f^2+ 5 m^2 c^2\right) - m^4 c^4 k_f^3 -3 m^6 c^6 k_f}{\sqrt{k_f^2 + m^2 c^2}} + m^6 c^6 \frac{\lambda^2 c^2}{2 E_p^2} \ln \left[ \frac{k_f + \sqrt{k_f^2 + m^2 c^2}}{m c} \right] \right. \\
     & \left. -\frac{8 \lambda c}{5 E_p} k_f^5 \right\}.   
\end{split}
 \label{eq22}
\end{equation}

\end{widetext}
As we can see, the expressions for energy density and pressure can  be divided in two terms such that in the limit $\lambda\rightarrow 0$, corresponding to the low energy limit, the second term in RHS vanishes. 
In what follows, we will numerically analyze in details these EoSs in the context of astrophysics of compact objects.

\section{TOV}\label{TOV}

\begin{figure*}[h]
\centering
\begin{tabular}{ll}
\includegraphics[width=6.2cm,angle=270]{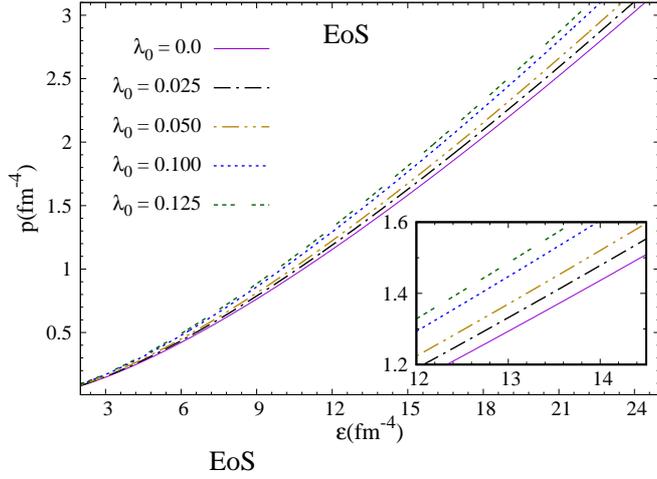}
\includegraphics[width=6.2cm,angle=270]{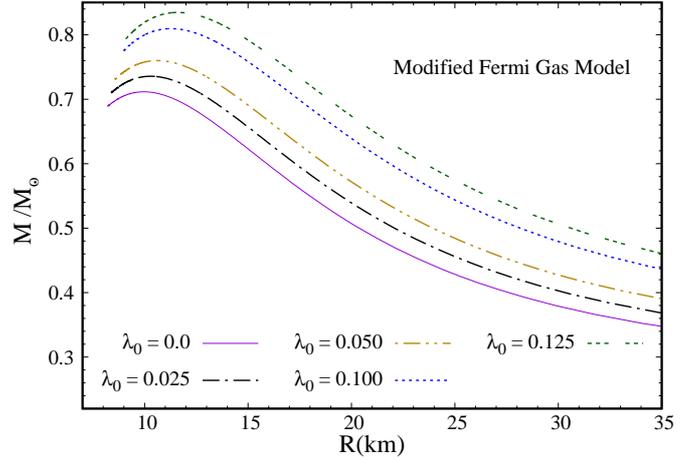}
\end{tabular}
\caption{Left: EoS for modified free Fermi gas in the first case. Right: mass and radius relation for a family of pure neutron stars. $\lambda_0 = 0$ recover the (Dirac) Fermi gas.}
\label{fig1}
\end{figure*}

\begin{figure*}[h]
\centering
\begin{tabular}{ll}
\includegraphics[width=6.2cm,angle=270]{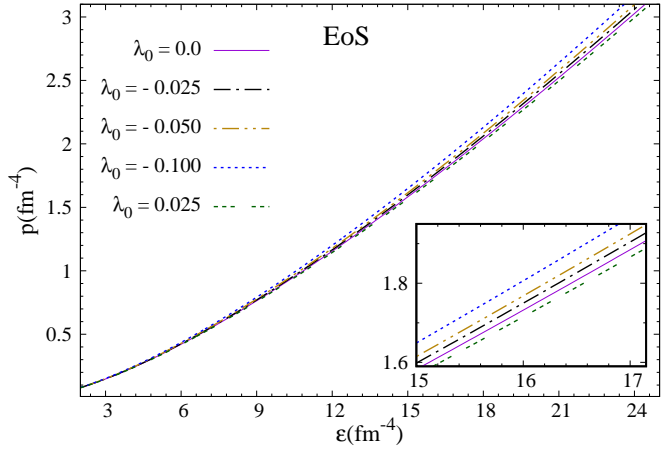}
\includegraphics[width=6.2cm,angle=270]{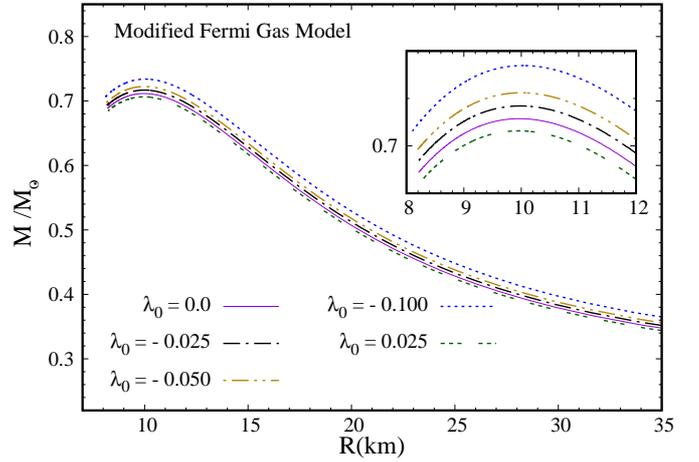}
\end{tabular}
\caption{Left: EoS for modified free Fermi gas in the second case. Right: mass and radius relation for a family of pure neutron stars. $\lambda_0 = 0$ recover the (Dirac) Fermi gas.}
\label{fig2}
\end{figure*}


In this section, we present the Tolman - Oppenheimer - Volkoff (TOV) equations. With the use of these equations, it is possible to analyze the effects of the  relations for the free Fermi gas when it is applied to the structure of compact stars.

Einstein's equations are the basis for the theory of General Relativity. The solutions of these field equations allow us to obtain a description of the hydrostatic equilibrium of isotropic, homogeneous, spherically symmetric and static (no rotation) objects \cite{Glendenning1997,Carroll2004}, a reasonable approximation for the so-called compact astrophysical objects, such as white dwarfs, magnetars, neutron stars and others. The birth of these objects coincides with the end of the nuclear fusion process of chemical elements inside them. The equilibrium conditions are guaranteed by the balance between the nuclear degeneracy pressure and the gravitational field. In agreement, the physics of these objects involves two steps: first, the construction of an EoS from nuclear physics; second, the use of relativistic hydrostatic equilibrium equations obtained from general relativity, the TOV equations.

For the description of these relativistic objects, Einstein's equations are carefully applied in the outer and inner regions of space-time. These regions being static and isotropic as a first approximation. The outer region is an asymptotically flat vacuum solution, given in terms of the Schwarzschild metric. These two parts must be carefully combined, with the stellar surface characterized by a point where the interior pressure vanishes. Einstein's equations for the stellar interior are given by: 
\begin{equation}
G_{\mu\nu} \doteqdot R_{\mu\nu}-\frac{1}{2}g_{\mu\nu}R = \frac{8\pi G}{c^4} T_{\mu\nu},
\end{equation}
where, $G_{\mu\nu}$ is the Einstein geometric tensor and $T_{\mu\nu}$ is the energy-momentum tensor, which is responsible for describing the matter in the stellar interior. We can model the matter of the star as a perfect fluid, therefore, $T_{\mu\nu}$ in a comoving frame is described as 

\begin{equation}
T_{\mu\nu}=(\varepsilon+p)u_{\mu}u_{\nu}-pg_{\mu\nu},
\end{equation}
where $p$ and $\varepsilon$ represents respectively the pressure and energy density of the fluid and are functions only of the radial coordinate $r$. The connection between $p$ and $\varepsilon$ is provided by means of an EoS. Note that we are using the metric signature $(+---)$. Thus, combining the results for $G_{\mu\nu}$ and for the tensor $T_{\mu\nu}$, we obtain two coupled differential equations: 
\begin{equation}
M'(r)=4\pi r^{2}\varepsilon(r),
\label{neweq1}
\end{equation}

\begin{equation}
        p'(r)
        = -\frac{GM(r)\varepsilon(r)}{r^{2}}\left(1+\frac{p(r)}{\varepsilon(r)c^2}\right) \left(1+\frac{4\pi r^{3} p(r)}{M(r)c^2} \right)\left(1-\frac{2GM(r)}{rc^2}\right)^{-1}.
     \label{neweq2}
\end{equation}
Finally, Equations (\ref{neweq1}) and (\ref{neweq2}) are the reduction of Einstein's equations to the interior of a static, isotropic and spherical relativistic compact star. These are the hydrostatic equilibrium equations for General Relativity, known as TOV \cite{tov1,tov2}. This set of coupled differential equations (\ref{neweq1}) and (\ref{neweq2}) enables us to describe some measurable macroscopic astrophysical properties of compact stars, such as mass and radius, analyzed in the next section. 

\section{Results and discussion}\label{results}

In this section we show the numerical results of the EoS and TOV equations shown in the previous sections. 

As discussed before, NSs are the final stage of the collapse of massive stars. During the dynamical process of collapse of these stars, nuclear processes occur, such as electron capture, which make the stellar matter richer in neutrons, in such a proportion that the proton fraction of the star is close to $Y_p = 0.1$ or less \cite{Glendenning1997} and most particles are neutrons. Therefore, in a good approximation, a NS is composed just by neutrons. However, it is known that in the NS we also have protons and electrons in beta-equilibrium with the neutrons, and in the inner core of the star more exotic states could appear, such as hyperons or quark matter \cite{Glendenning1997}. 
Since the purpose of this work is to analyse the modifications due to deformed fermionic kinematics, a single species is the simpler situation. Therefore we continue our analysis applying the DSR equations to pure neutron matter.

Before discussing our astrophysical relevant results, we would like to comment that for  reasons of simplicity, at this stage, $\lambda$  is  redefined as $\lambda = \frac{\lambda_0 E_{p}}{c^2 m}$. In the following, we analyze the values for $\lambda_0$.

In the left panel of  Fig. \ref{fig1} we show the evolution of the pressure with respect to energy density for the first case (Eq. (\ref{eq1})) of the modified Fermi gas. Note that $\lambda_0 = 0$ recovers the Fermi gas within special relativity. We increase the parameter $\lambda_0$  from $0.025$ to $0.125$ and note that, for a given energy density, the pressure increases along with increasing values of $\lambda_0$, {\it i.e.}, the modified Fermi gas has a stiffer EoS compared to the special relativity.
The consequence of the stiffer EoS can be seen in the mass $\times$ radius diagram of a family of NS (right panel of Fig. \ref{fig1}) constructed with only kinetic contribution of pure neutron matter. We note that the modified special relativity predicts mass and radius higher than the special relativity for $\lambda_0>0$. 

If $\lambda>0$, from the dependence of the energy with the momentum and mass in the modified dispersion relation, we verify a decrease in the energy for given values of momentum and mass in first order perturbation. Such reduction is fixed, since it does not depend on the momentum, but just on the mass of the particle, as can be seen from the term $A_1$ of Eq.(\ref{eq8}). Physically one could interpret that a portion of the particle’s energy is reduced due to the interaction with degrees of freedom of the quantum spacetime described phenomenologically by the dispersion relation. On the other hand, a feature of this relation is the fact that its group velocity, $dE/dk$, is undeformed in first order perturbation in the Planck scale (corrections start at second order), i.e., it does not follow the reduction verified for the energy, or at least it reduces much less when analyzing higher order corrections. For this reason, for each value of pressure (as it depends on the group velocity from Eq.(8)), one attributes a corresponding smaller value of energy. Therefore, we are verifying, in fact, a relative increment of the group velocity in comparison to the particle's energy, and this features physically manifests as a stiffness of the EoS when we compare the pressure versus the energy density. The opposite behavior should happen if $\lambda <0$.

Finally, in Fig. \ref{fig2} we analyze the effects for the second case corresponding to the set of functions of Eq. \ref{eq2}. Notice that for $\lambda_0=0$ the conclusions are the same as discussed above. In this case, we test negative and positive values for $\lambda_0$, since in this modification, negative values of $\lambda_0$ are the ones that yield stiffer EoS and bigger maximum mass. And unlike the previous case, we note that positive values of $\lambda_0$ reproduce softer EoS, resulting in mass-radius relation with slightly smaller maximum mass values when compared to Fermi gas in special relativity. Therefore we decided, for reasons of comparison, to show only one positive value of this parameter, $\lambda_0=0.025$.

From a perturbative analysis, if $\lambda>0$ we verify a momentum-dependent decrease in the energy at first order, (as can be seen in the last term of the RHS of Eq.(\ref{eq19})), which implies in a decrease in the group velocity at the same order. Considering this ``competition'' between the fermion energy and the group velocity due to Planck scale effects, we verify a stronger reduction in the latter (taking into account the full integrand of Eq.(\ref{eq7})) than the former. For this reason, analyzing in a relative way, the fluid manifests as being less (more) stiff for positive (negative) values of $\lambda$. In summary, the stiffness of the corrected fermion gas depends on the competition between the corrections on the particles’ energies and the group velocity measured by the modified dispersion relation.

\section{Final Remarks}
\label{conclusion}

In this paper, we use DSR to obtain the EoS that describes a Fermi gas whose constituents obey a modified dispersion relation. We derive the new expressions for the pressure and energy density within two cases.
 The first one is given by Eqs. (\ref{eq1}) and predicts a stiffer EoS for positive values of $\lambda_0$. Furthermore, the exponential function, Eq. (\ref{eq2}), of the second case predicts stiffer EoS for negative values of the DSR parameter.

Given that NS are mostly composed by neutrons, we use a simplified version of NS that are composed by non-interacting neutrons. We note that in a more realistic model for NS in nature, it is necessary a nuclear interaction, and the presence of other particles, {\it e.g.}, protons and electrons in beta-equilibrium with the neutrons.
 The simplified model presented in this work brings clear signs of the effect of modified special relativity on a free Fermi gas, which is a modification of the stiffness of the EoS compared to the special relativity one.  As a future work, it will be interesting to extend our approach to study an EoS with modified relation dispersion, including nuclear interaction applied to TOV, in order to understand the effect of DSR in more realistic models. Another improvement for a future work is the implementation of DSR at finite temperature, which is important if one is interested to study EoS for supernovae for example. In such case, the Fermi distribution present on Eq. \eqref{eq4} of our manuscript will be taken with $T > 0$.  We expect the effect of temperature on the EoS to be similar of the one in special relativity, when, at a given density, the pressure increases due to the thermal energy.

The two main aspects of the present work are: (i) while most of current works on EoS investigates modifications in the nuclear force, we approach the kinetic contribution of the nuclear EoS, which is usually taken either by a non-relativistic Fermi gas or by a relativistic one described by special relativity; (ii) we note that DSR predicts stiffer EoS and higher NS maximum mass depending on the phenomenological models.

\section*{Acknowledgements}
 CEM has a scholarship paid by Capes (Brazil) and LCNS would like to thank Conselho Nacional de Desenvolvimento Cient\'ifico e Tecnol\'ogico (CNPq) for partial finnancial support through the research Project No. 164762/2020-5. IPL would like to acknowledge the contribution of the COST Action CA18108 and National Council for Scientific and Technological Development - CNPq grant 306414/2020-1.

\bibliographystyle{ieeetr}
\bibliography{referencias.bib}

\end{document}